\documentclass[11pt]{article}

\usepackage{url}
\usepackage{alltt}
\usepackage{amssymb}

\usepackage{epsfig}
\usepackage{wrapfig}
\usepackage{subfigure}

\usepackage{rotating}
\usepackage{moreverb}
\usepackage{fancyvrb}

\def\bd{\begin{description}}
\def\ed{\end{description}}
\def\bc{\begin{center}}
\def\ec{\end{center}}
\def\bq{\begin{quote}}
\def\eq{\end{quote}}
\def\bi{\begin{itemize}}
\def\ei{\end{itemize}}
\def\be{\begin{enumerate}}
\def\ee{\end{enumerate}}
\def\ba{\begin{array}}
\def\ea{\end{array}}

\newcommand{\lv}[1]{ }

\begin{document}

\title{Rule-Based Drawing, Analysis and Generation of Graphs for Mason's Mark Design\\
{\normalsize Tool Description}}  
\author{%
Thom Fr{\"u}hwirth\\
Ulm University, Germany\\
\url{http://www.constraint-handling-rules.org}  
}

\maketitle

\begin{abstract}
We are developing a rule-based implementation of a tool to analyse and generate graphs. 
It is currently used in the domain of mason's marks. 
For thousands of years, stonemasons have been inscribing these symbolic signs on dressed stone.
Geometrically, mason's marks are line drawings. 
They consist of a pattern of straight lines, sometimes circles and arcs.
We represent mason's marks by connected planar graphs.

Our prototype tool for analysis and generation of graphs is written in the rule-based declarative language Constraint Handling Rules. 
It features
\begin{itemize}
\item a vertex-centric logical graph representation as constraints,
\item derivation of properties and statistics from graphs,
\item recognition of (sub)graphs and patterns in a graph,
\item automatic generation of graphs from given constrained subgraphs,
\item drawing graphs by visualization using svg graphics
\end{itemize}
In particular, we started to use the tool to classify and to invent mason's marks. 
In principe, our tool can be applied to any problem domain that admits a modeling as graphs.

\end{abstract}

\section{Introduction}

Mason's marks are symbols often found on dressed stone in historic buildings.
These signs go back about 4500 years, to the tombs of the advanced ancient civilization of
Egypt. 
In Europe, 
they were common from the 12th century on \cite{friedrich1932,davis1954catalogue}. 
There, one can mainly find mason's marks from the medieval ages, 
mostly in churches,  cathedrals and monasteries. 
In one such building, there may be a thousand mason's marks of hundred different designs.
Over time, mason's marks got smaller and more complex.  

These mason's marks were inscribed on the stones by stonemasons during construction of a building to identify their work, presumably 
for quality control and probably to receive payment. 
This was important for masons who were free of servitude and therefore allowed to travel the country for work (see \cite{follett2010pillars} for popular fiction on the subject). 
Only certain stonemason masters were allowed to inscribe their mark in a blazon, like a signature. 
Their mason's marks are also found on medieval documents and masons tombstones.
The master would give a personal mason's mark to his apprentices, provided they had enough skill to construct and interpret the mason's mark symbol. A challenging design and interpretation would make it harder to appropriate or misuse a mason's mark. 
Stonemason's mark are an important source of information for art and architecture historians and archaeologists, 
in particular to reconstruct the construction process of buildings.

Mason's marks tend to be simple geometric symbols, usually constructed using rulers and compasses and precisely cut with a chisel. In this way a distinctive sign consisting of straight lines and curves could be produced  with little effort.
The geometric construction of mason's marks implies that they exhibit a structural regularity. 
This was first discussed and formalized by Franz von R{\v{z}}iha \cite{rziha1881}. He claimed that mason's marks are small subfigures of regular grids, which consist either of squares or equilateral triangles together with circles inscribed in each other.
His theory is obsolete, because there is no historical evidence that grids were explicitly used and because not all marks fit these grid patterns. 
It is however clear that the construction of the mason's marks with compasses and rulers
lends itself to the prevalence of certain angles 
(multiples of $30$ and $45$ degrees)
between lines and certain multiples of line lengths 
($1, \sqrt{2}, \sqrt{3}, 2, \sqrt{5}\ldots$).

We are developing a graph tool and apply it to draw, analyse and generate mason's mark designs.
In this paper, we shortly present the representation of straight-line graphs, 
how for analysis we produce statistics about graph properties, and how we recognize subgraphs using pattern matching rules.
Then we introduce our node-centric representation of line graphs and describe how to exhaustively or randomly generate graphs from given small constrained subgraphs.
We conclude by discussing related and future work.

\section{Tool Description}

Our prototype graph analysis and generation tool is currently implemented using CHR in SWI Prolog \cite{swimanual}. We assume some basic familiarity with Prolog and Constraint Handling Rules (CHR)
\cite{fruhwirth2015constraint,chr-book}. 
Constraints are special relations, predicates of first order logic.
They are defined by CHR rules, consisting of a left-hand-side, a guard and a right-hand-side. 
When the constraint pattern on the left-hand-side matches some of the current constraints and the guard test (precondition) succeeds under this matching, the right hand side of the rule is executed. Depending on the rule type, matched constraints may be removed or not. The right hand side will typically compute and add new constraints.

\subsection{Representation of Mason Marks as Graphs}

We represent mason's marks (without arcs) by connected 
{\em planar straight-line graphs}, a
drawing of planar graphs in the plane such that its edges are straight line segments
\cite{tamassia2013handbook}.
A line (segment) has two nodes, or endpoints, given by Cartesian coordinates.
Each point is defined by a pair of numbers written {\tt X-Y}. 
For convenience of manipulation, we redundantly represent lines at the same time by polar coordinates, which consist of a reference point (pole), which is the first endpoint of the line, a line length (radius) and an angle (azimuth) in degrees. 
This leads to the line constraint:
\begin{verbatim}
 l(EndPoint1, EndPoint2, LineLength, Angle)
\end{verbatim} 
With polar coordinations, translation, rotation and scaling of lines is straightforward. 
Only when the lines are drawn, missing point coordinates are computed:
\begin{verbatim}
l(X1-Y1,P2,L,A) ==> numbers(X1,X2,L,A) | 
           X2 is X1+L*U*cos(A*pi/180), Y2 is Y1+L*U*sin(A*pi/180),
           P2=(X2-Y2).
 % analogously for point P1 when P2 is known
\end{verbatim} 
This {\em propagation rule} of the form {\tt LeftSide ==> Guard | RightSide} can be read as follows:
If a line matching {\tt l(X1-Y1,P2,L,A)} is found where {\tt X1,X2,L,A} are numbers (instead of yet unbound variables), then compute {\tt X2} and {\tt Y2} using Prolog's {\tt is} built-in and equate (unify) the endpoint {\tt P2} with the coordinate {\tt X2-Y2} using Prolog's {\tt =} built-in. 
If {\tt P2} was a free (unbound) variable, it will be bound, otherwise a equality check will be performed.

\subsection{Analysis of Graphs}

From a given graph, i.e. its line constraints, we can generate information using propagation rules. 
For example, one can compute counts for the occurrences of each value in the components of a line (points, lengths, angles) to collect statistical information about the graph. 
Note that the number of occurrences of a node corresponds to the degree of that node.

The constraint {\tt a(Type, Count, Value)} can be considered as an array entry that contains for each {\tt Value} of a certain {\tt Type} its {\tt Count} of occurrences. Below, the first rule adds such entries for the same {\tt Type,Value} pair.
The second rule computes relevant information from a single line.
\begin{verbatim}
 % add counts for two entries of the same T(ype), V(alue) pair
a(T,N1,V), a(T,N2,V) <=> N is N1+N2, a(T,N,V). 

 % compute statistical information about lines of a graph
 % Types: l(ine)c(ount), n(ode), l(ine )l(ength), a(ngle)
l(P1,P2,L,A) ==> a(cl,1,l),a(n,1,P1),a(n,1,P2),a(ll,1,L),a(a,1,A).
\end{verbatim}
The first rule is a {\em simplification rule} without a guard. It replaces two matching {\tt a} constraints by a new one containing the sum.
We can compute relative angles and line length proportions between lines as follows.
\begin{verbatim}
 % angles between lines that share a node, e.g. first node
l(P1,P2,L1,A1), l(P1,P4,L2,A2) ==> A is abs(A1-A2), a(al,1,A).

 % proportions between lines lengths of any two lines in a graph
l(P1,P2,L1,A1), l(P3,P4,L2,A2) ==> R is L1/L2, a(pl,1,R).
\end{verbatim}
Any other type of information can be derived from the lines of a graph as well.

\subsection{Pattern Matching of Graphs}

We want to find patterns and recognize subgraphs in a graph.
For recognition we assume that all lines have angles between $0$ and $180$ degrees. 
(Lines with an angle between $180$ and $360$ degrees can be inverted.)
Note that that size and orientation of graphs can differ. 
To account for scaling and rotation, we introduce two Prolog predicates that we will use in the guard of rules. 

The predicate {\tt proportional(Ls,Ps,R)} accounts for scaling. It checks that the ratio between the next element from the list {\tt Ls} and the next element from the list {\tt Ps} is always the ratio {\tt R}. The intended use is that {\tt Ls} is a list of actual line lengths while {\tt Ps} is a list of required proportions between these line lengths. 
Analogously, the predicate {\tt rotated(Ls,Ps,A)} accounts for rotation. 
\lv{
It checks that all angles in list {\tt Ls} can be obtained from the angles in list {\tt Ps} by a rotation by angle {\tt A}.
}
\begin{verbatim}
proportional([],[],R).
proportional([L|Ls],[P|Ps],R):- R is L/P, proportional(Ls,Ps,R).

rotated([],[],A).
rotated([L|Ls],[P|Ps],A):- A is (L-P) mod 360, rotated(Ls,Ps,A).
\end{verbatim}
The above code can be modified to account for imprecisions in the measurements of lengths and angles by using rounding or interval arithmetics. 

\begin{figure}[ht]
\centering
\begin{minipage}{.22\textwidth}
	\centering
  \includegraphics[width=\textwidth]{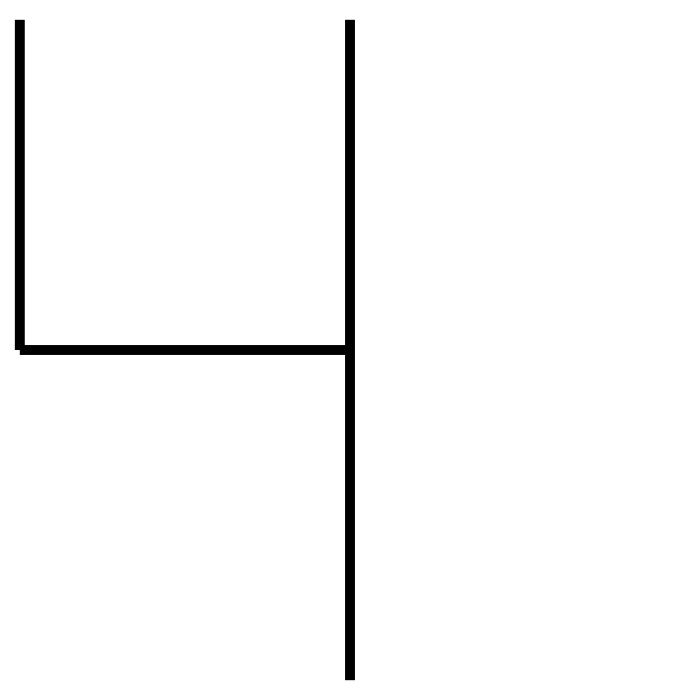}  
	\caption{Y-Sign Graph}
	\label{y_sign}
\end{minipage}%
\begin{minipage}{.02\textwidth}
  \ \ \
\end{minipage}%
\begin{minipage}{.22\textwidth}
	\centering
  \includegraphics[width=\textwidth, angle=90]{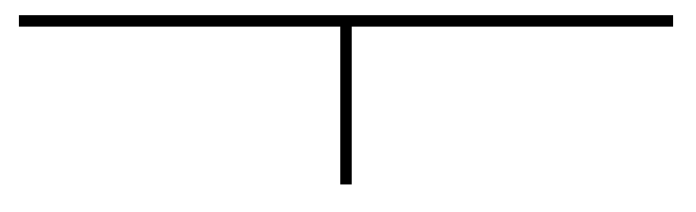}  
	\caption{Graph {\tt [2,90,1,90,2]}}
	\label{t_sign}
\end{minipage}%
\begin{minipage}{.03\textwidth}
  \ \ \
\end{minipage}%
\begin{minipage}{.22\textwidth}
	\centering
  \includegraphics[width=\textwidth]{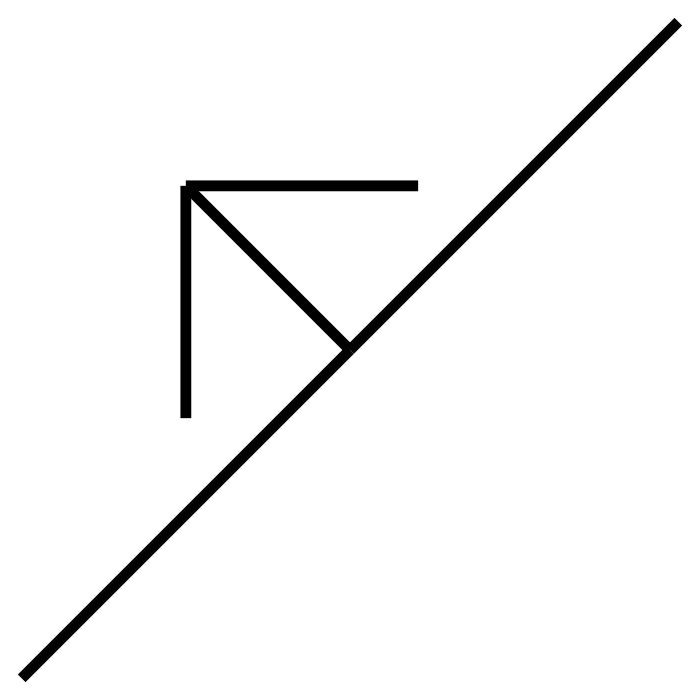}  
\caption{Graph {\tt [2,90,1-I,90,2], [3,45,3-I,45,3]}}
	\label{mark1}
\end{minipage}
\begin{minipage}{.03\textwidth}
  \ \ \
\end{minipage}%
\begin{minipage}{.22\textwidth}
	\centering
  \includegraphics[width=\textwidth, angle=90]{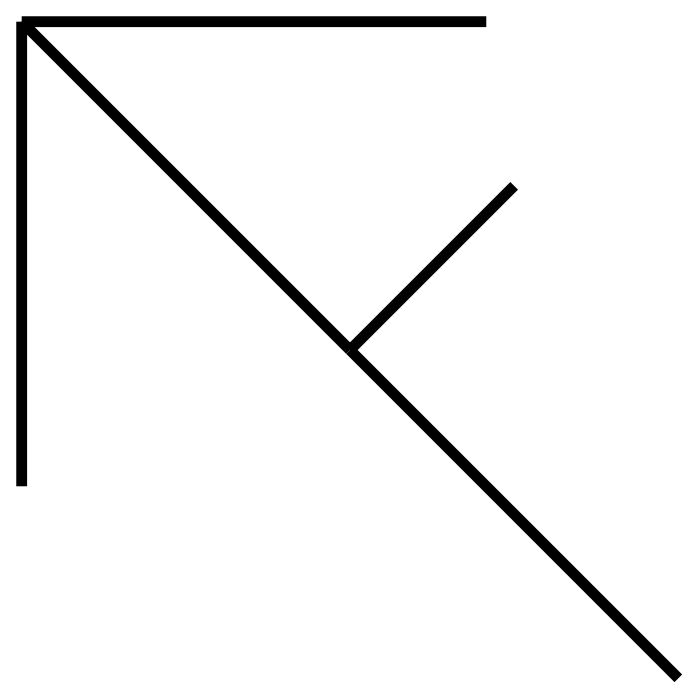}  
\caption{Graph {\tt [2-I,90,1,90,2], [3,45,3-I,45,3]}}
	\label{mark2}
\end{minipage}
\end{figure}

Below are two examples: how to recognize parallel lines and the subgraph depicted in Figure \ref{y_sign}.
We use a constraint {\tt recognized(What,NodeList)} to record what has been recognized for which nodes.
\begin{verbatim}
 % two parallel lines have the same angle
l(A,B,L1,A), l(C,D,L2,A) ==> recognized(parallel,[A,B,C,D]).

 % recognize subgraph comprised of four lines given in Figure 1
l(A,B,L1,A1), l(B,C,L2,A2), l(E,C,L3,A3), l(C,D,L4,A4) ==> 
                rotated([A1,A2,A3,A4],[90,0,90,90],_), 
                proportional([L1,L2,L3,L4],[1,1,1,1],_) | 
                                 recognized(y_sign,[A,B,C,D,E]).
\end{verbatim}

\subsection{Node-Centric Representation of Graphs Using Half-Lines}

Through exhaustive initial experiments we found that for the encoding and generation of mason's marks a {\em node-centric} (vertex-centric) representation of their underlying connected straight-line graph is helpful. The constraint for a node is defined as follows:  
\begin{verbatim}
 node(NodePoint, NodeList)   
 NodePoint ::= Number-Number
 NodeList  ::= [LineLength] ; [LineLength,Angle|NodeList]
\end{verbatim}
Each node is at the center of several lines leaving it. We record the length of these lines and the angle between neighboring lines. 
{\tt NodeList} is a list with elements that are alternating between line lengths and angles, starting and ending in a line length.
We may omit the {\tt NodePoint} and just use {\tt node(NodeList)}.
For example {\tt node([2,90,1,90,2])} depicts the graph given in Figure \ref{t_sign}. 
Note that each such node forms a small subgraph by itself.

In order to describe larger connected graphs,
the {\tt node} data structure representation is extended to allow identifiers for lines. 
These identifiers are optionally attached to the line-lengths.
If such an identifier is shared between two lines in different nodes it means that these lines are the same, with the two nodes as endpoints. 
Such annotated lines we call {\em half-lines}, because one needs a matching pair of them to form a valid line.

When the nodes are translated into line constraints,
the subgraphs of the two nodes connected by this common line are scaled and rotated such that the half-lines become identical.
For example {\tt node([2,90,1-I,90,2]),} {\tt node([3,45,3-I,
45,3])} depicts the graph given in Fig. \ref{mark1}. In effect, the subgraph {\tt node([3,45,3-I,
45,3])} is rotated and scaled to meet the half-line in {\tt node([1,45,1-I,45,1])}. Contrast this with the situation in Figure \ref{mark2}, where the half line of the first node has changed.

\subsection{Exhaustive Generation of Arbitrary Graphs}

We can exhaustively generate graphs from a given node-centric representation containing half-lines with free unbound logical variables as identifiers. 
Different resulting graphs are possible, depending on which half-lines are identified
by binding (aliasing) their variables.
Not all such matchings lead to a valid graph, 
because the resulting graph may not be geometrically possible.

In our implementation, the given graph in node-centric representation
is translated into a conjunction of lines (some of them are half-lines).
First, all identifiers of half-lines are collected into a list.
The identifiers can be unbound variables, 
in that case identifiers can be bound to each other, making them equivalent.
Then a recursion on this list using Prolog predicate {\tt pairlines} 
equates the next identifier in the list with one of the remaining identifiers
using the Prolog built-in {\tt select(Element,List,RestList)}
that non-deterministically removes an element from a list.
Recursion continues with the remaining list {\tt L1}. 
 %
\begin{verbatim}
pairlines([I|L]) :- select(I,L,L1), pairlines(L1).      
pairlines([]).          
\end{verbatim}

Two half-lines with identical identifiers react with each other by following rule.
The purpose of the rule is to connect the two subgraphs in which the 
two half-lines occur by merging the two lines into one. 
Essentially the same non-deterministic paring rules were used in different application domain in \cite{fruhwirth1998optimal}.
Scaling and rotation is applied to the complete subgraph in which the second half-line occurs using the constraint {\tt update(Node,Scaling,Rotation)}.
Finally, the nodes of the two half lines are identified and a new proper full line without identifier replaces the two merged half-lines. (At this point, the node points are still unbound variables, their coordinates have no values yet.)
\begin{verbatim}
 % find two half-lines whose half-line identifier is the same
 % to connect their subgraphs
l(N1,N2,M1-I,A1), l(N3,N4,M2-I,A2) <=> 
    alldiff([N1,N2,N3,N4]),     % all nodes must be different
    M is M1/M2, A is A1+180-A2, % compute scaling and rotation
    update(N3,M,A), % scale and rotate N3 graph to fit N1 graph
    N1=N4, N2=N3,   % equate nodes of now identical half-lines
    l(N1,N2,M1,A1). % merged line replaces the two half-lines
\end{verbatim}
 
The rule for updating a line by constraint {\tt update(Node,Scaling,Rotation)}
using scaling and rotation and for updating all lines connected to that line is shown below.
It has to avoid repeated updates of the same line. Therefore the update removes the line and produces an intermediate representation of the line using constraint {\tt lu}. 
\begin{verbatim}
 % update line with node N1 by scaling and rotation
update(N1,M,A) \ l(N1,N2,M1,A1) <=>  numbers(M1,A1) | 
     M2 is M1*M, A2 is A+A1, 
     lu(N1,N2,M2,A2),  % intermediate updated line
     update(N2,M,A).   % propagate update to node N2
 % same for line with node N2 to update
\end{verbatim}
This {\em simpagation rule} keeps the {\tt update} constraint but removes the line constraint {\tt l}.
It is replaced by the updated line constraint {\tt lu} and the update is also applied to node {\tt N1}.
Only when all updates are done, all these intermediate {\tt lu} lines are replaced by original {\tt l} lines with the help of some additional rules.

\subsection{Random Generation of Graphs for Mason Marks}

\begin{figure}[htb]
\centering
\begin{minipage}{.2\textwidth}
	\centering
  \includegraphics[width=\textwidth]{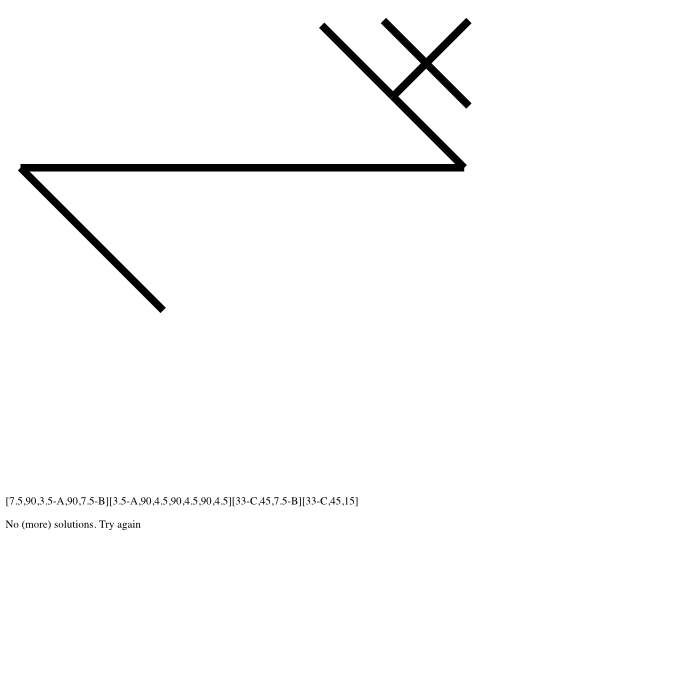}  
\end{minipage}%
\begin{minipage}{.2\textwidth}
	\centering
  \includegraphics[width=\textwidth]{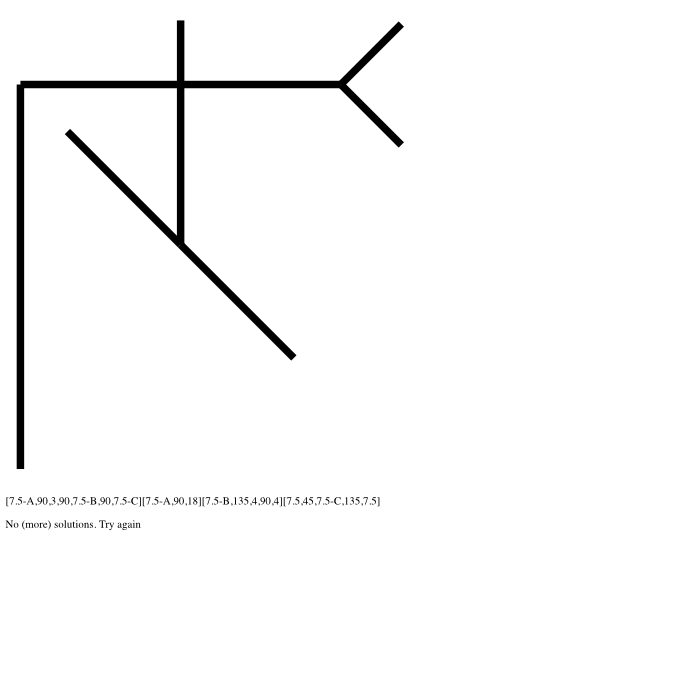}  
\end{minipage}%
\begin{minipage}{.2\textwidth}
	\centering
  \includegraphics[width=\textwidth]{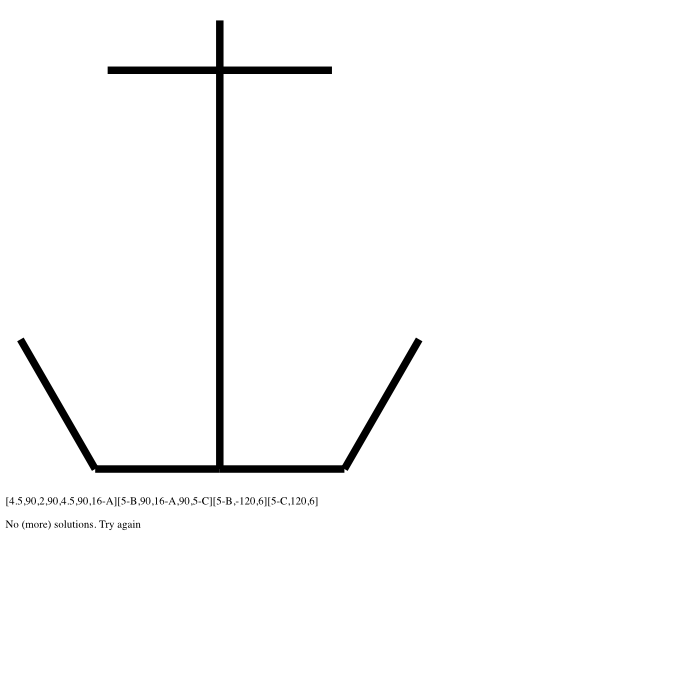}  
\end{minipage}%
\begin{minipage}{.2\textwidth}
	\centering
  \includegraphics[width=\textwidth]{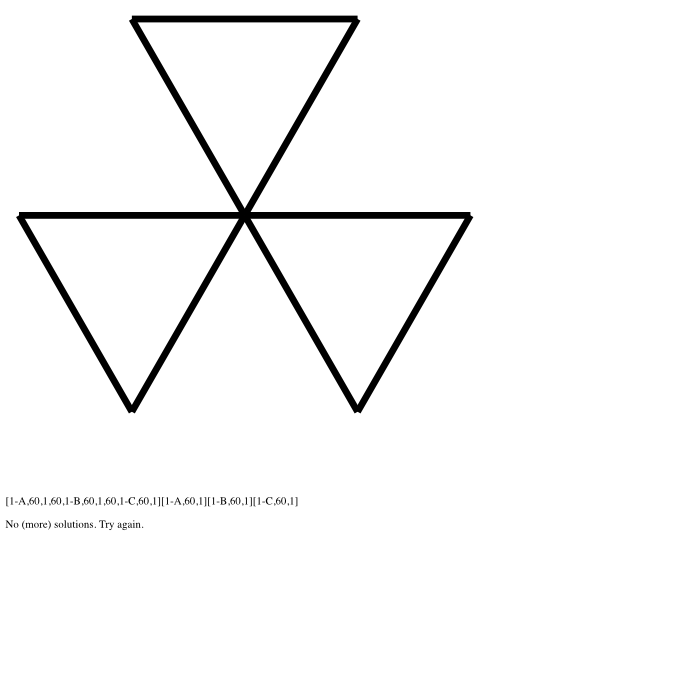}  
\end{minipage}%
\begin{minipage}{.2\textwidth}
	\centering
  \includegraphics[width=\textwidth]{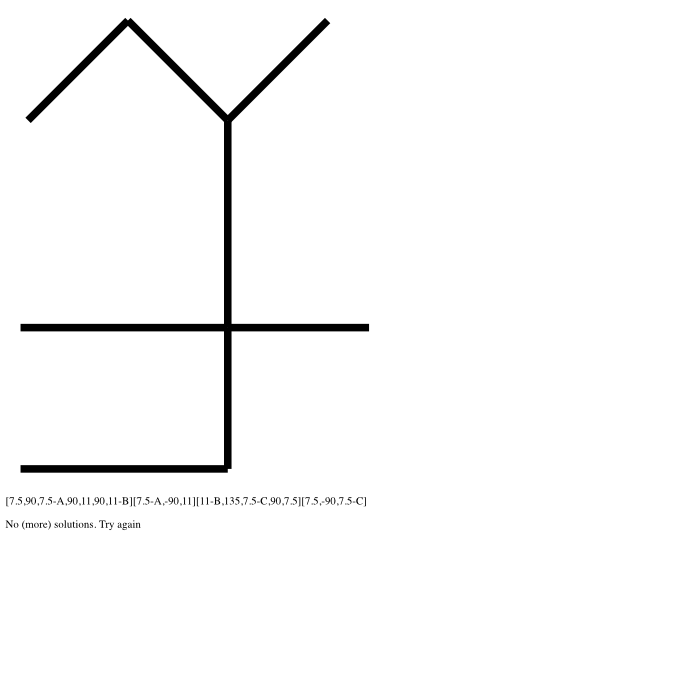}  
\end{minipage}%
	\caption{Mason's Marks of Ulm Minster}
	\label{ulm}
\end{figure}

We have encoded a number of mason's marks from \cite{rziha1881} in our node-centric representation,
in particular for the Ulm Minster (see Figure \ref{ulm}).  
The minster has thousands of marks of about 120 different designs.
Most of these marks can be described using just 4 to 5 {\tt node} constraints in our node-centric representation.
In each mason's mark there is typically a node that is connected to most nodes and that is located near the geometric center of the mason's mark. Such a node is heuristically chosen as the 
{\em primary node} of the mason's mark graph. 
We collected all {\tt node} constraints for primary nodes and for all other nodes from our encoding of existing mason's marks. 

\begin{figure}[htb]
\centering
\begin{minipage}{.2\textwidth}
	\centering
  \includegraphics[width=\textwidth]{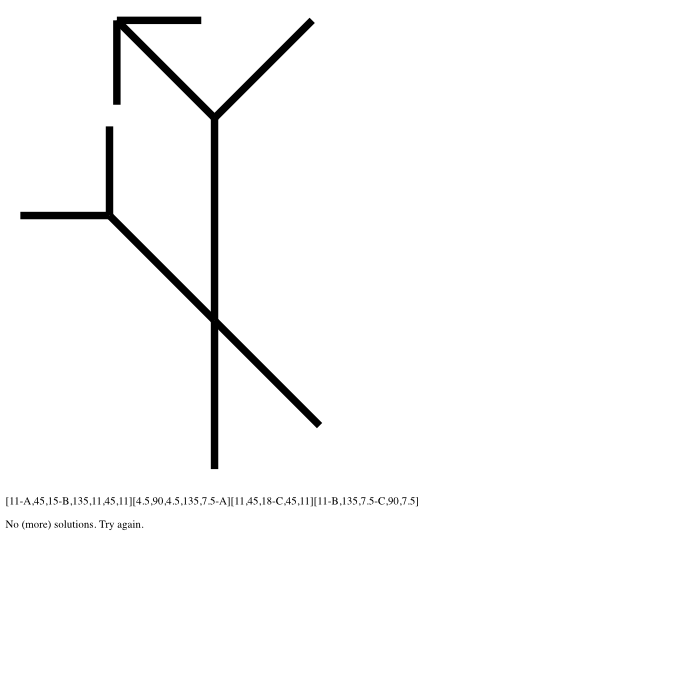}  
\end{minipage}%
\begin{minipage}{.2\textwidth}
	\centering
  \includegraphics[width=\textwidth]{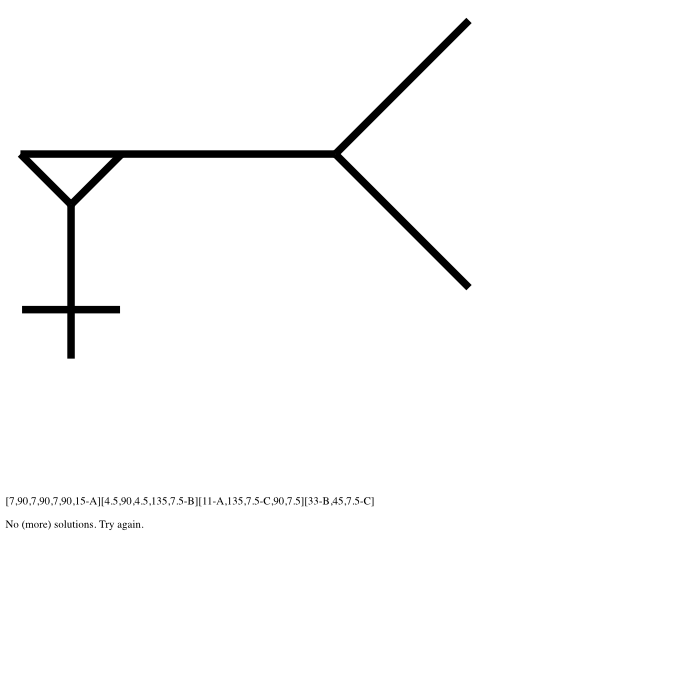}  
\end{minipage}%
\begin{minipage}{.2\textwidth}
	\centering
  \includegraphics[width=\textwidth]{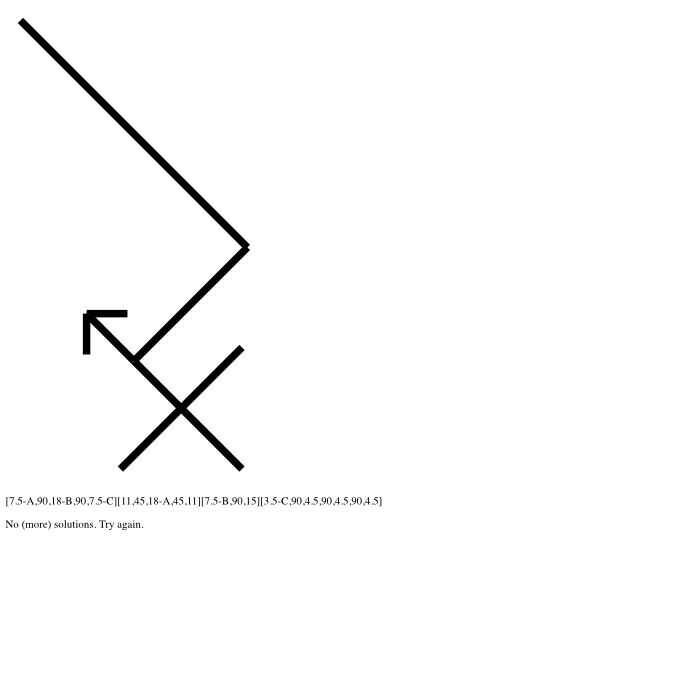}  
\end{minipage}%
\begin{minipage}{.2\textwidth}
	\centering
  \includegraphics[width=\textwidth]{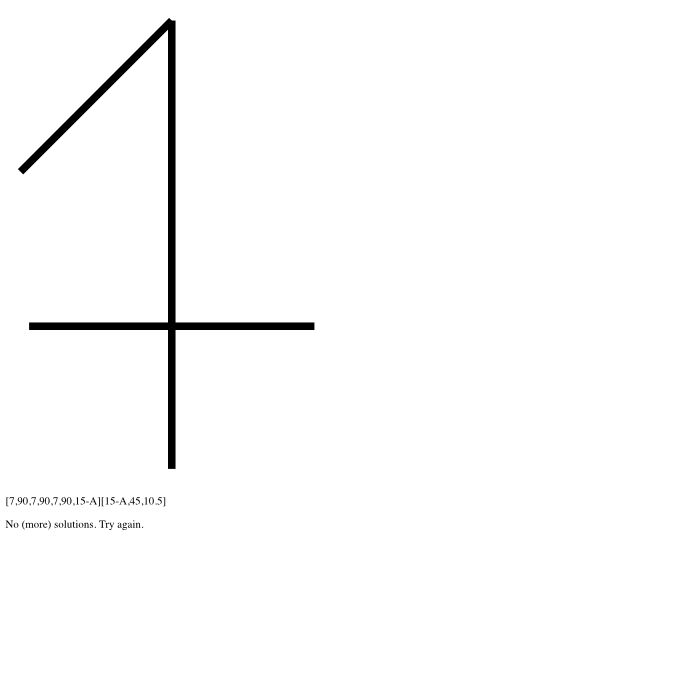}  
\end{minipage}%
\begin{minipage}{.2\textwidth}
	\centering
  \includegraphics[width=\textwidth]{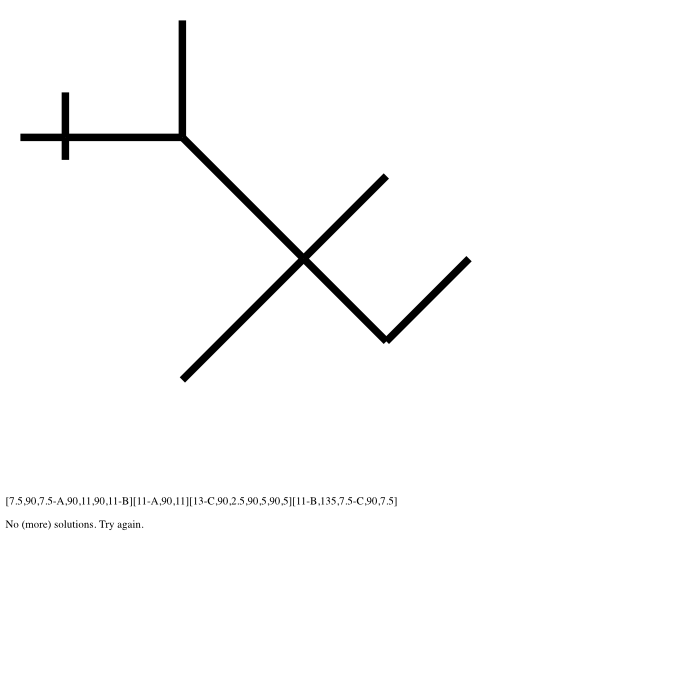}  
\end{minipage}%
	\caption{Randomly Generated Mason's Marks Derived from Ulm Minster Marks}
	\label{random}
\end{figure}

For random generation of similarly shaped mason's mark graphs, first we choose one primary node and two other nodes randomly, and then check if their angles are either multiples of 30 or 45 degrees.
If so, we choose a fourth node from the remaining other nodes (ignoring duplicates).
The resulting graph may not be valid due to unmatched half-lines. 
We then use Prolog's backtracking to try all possible fourth nodes. 
In this way, zero, one or more valid mason's marks are randomly produced
from the given sub-graph nodes.
Figure \ref{random} shows some examples of mason's marks generated in this way. 

\section{Related Work}

To the best of our knowledge our work is the first that not only represents, 
but also analyses and generates mason's marks. 
The only related work we could find is \cite{kiiko2002recognition}. 
Given a prototype graph of a mason's mark and the skeleton graph of an input image, the recognition of this image is considered as search of matchings paths in the skeleton graph
with a minimal number of mismatchings.
It is not clear from the description of the algorithm how it accounts for scaling and rotation.
In contrast, our recognition rules match line edges directly, independent of scale and rotation. 
Efficiency of matching relies on optimizations of the CHR compiler such as {indexing}. 
Imprecision could be taken into account and controlled by rounding the numerical values of coordinates, lengths and angles
or by using interval arithmetics.

In the work \cite{durst1996prolog} structural character descriptions for East
Asian ideograms (Kanji font characters) are analyzed and generated.  Sketches of
characters are produced from a symbolic coordinate free
description, which is a system of constraints.  
The authors developed a special finite domain constraint
solving algorithm tailored to the problem in CHR.
This approach proved to be more efficient and versatile than using existing built-in solvers.
Kanji characters are decomposed into subfigures and those are described by strokes (lines) called bars. The representation of bars is similar to our node-centric representation. However, their work employs only the four main directions and lengths are always implicit, while we allow for arbitrary angles and arbitrary explicit lengths.

\section{Conclusions and Future Work}

We have shortly presented our prototype tool to analyse, generate and draw straight-line graphs based on a novel node-centric representation of graphs using constraints. 
We have applied the tool to the domain of stonemason's marks.
For a more complete coverage of mason's marks, we need to add the representation of arcs and other curves.
We currently work on representing the mason's marks found on buildings in the Alicante province in Spain,
because these have not been encoded so far.

There is a similarity between our node-centric approach for mason's marks and the representation of chemical structures using molecules with their bindings. We therefore think that our approach can readily be applied to model and analyse chemical compounds as straight-line graphs. 
Another application area might be the visualization of temporal constraint networks \cite{fru_temporal_reasoning_techrep94}.

\bibliographystyle{alpha} 
\bibliography{masonmarks}

\label{lastpage}

\end{document}